\documentclass[12pt,preprint]{aastex}
\usepackage{amsmath,amssymb}
\usepackage{geometry}
\usepackage{graphicx}
\usepackage{subfigure}

\newcommand{\ha}{H$\alpha~$}
\newcommand{\hb}{H$\beta~$}
\newcommand{\oiii}{[O {\footnotesize III}](5007\AA)}
\newcommand{\sii}{[S {\footnotesize II}](6716\AA+6731\AA)}
\newcommand{\Zsolar}{$Z_{\odot}~$}
\newcommand{\msp}{$~$}

\slugcomment{}

\shorttitle{m83}
\shortauthors{Hong}

\begin{document}


\title{Large-scale shock-ionized and photo-ionized gas in M83: the impact of star formation}

\author{ 
Sungryong Hong\altaffilmark{1},
Daniela Calzetti\altaffilmark{1},
Michael A.~Dopita\altaffilmark{2},
William P.~Blair\altaffilmark{3},
Bradley C.~Whitmore\altaffilmark{4},
Bruce Balick\altaffilmark{5},
Howard E.~Bond\altaffilmark{4},
Marcella Carollo\altaffilmark{6},
Michael J.~Disney\altaffilmark{7},
Jay A.~Frogel\altaffilmark{8},
Donald Hall\altaffilmark{9}, 
Jon A.~Holtzman\altaffilmark{10},
Randy A.~Kimble\altaffilmark{11},
Patrick J.~McCarthy\altaffilmark{12},
Robert W.~O'Connell\altaffilmark{13},
Francesco Paresce\altaffilmark{14},
Abhijit Saha\altaffilmark{15},
Joseph I.~Silk\altaffilmark{16},
John T.~Trauger\altaffilmark{17},
Alistair R.~Walker\altaffilmark{18},
Rogier A.~Windhorst\altaffilmark{19}, 
Erick T.~Young \altaffilmark{20}
and 
Max Mutchler \altaffilmark{4}
}

\altaffiltext{1}{Department of Astronomy, University of Massachusetts,  
Amherst, MA 01003}
\altaffiltext{2}{Research School of Astronomy \& Astrophysics,  The  
Australian National University, ACT 2611, Australia}
\altaffiltext{3}{Johns Hopkins University, Baltimore, MD, USA}
\altaffiltext{4}{Space Telescope Science Institute, Baltimore, MD 21218}
\altaffiltext{5}{Department of Astronomy, University of Washington, 
Seattle, WA 98195-1580}
\altaffiltext{6}{Department of Physics, ETH-Zurich, Zurich, 8093  
Switzerland}
\altaffiltext{7}{School of Physics and Astronomy, Cardiff University, 
Cardiff CF24 3AA, United Kingdom}
\altaffiltext{8}{Association of Universities for Research in
Astronomy, Washington, DC 20005}
\altaffiltext{9}{Institute for Astronomy, University of Hawaii,
Honolulu, HI 96822}
\altaffiltext{10}{Department of Astronomy, New Mexico State
University, Las Cruces, NM 88003}
\altaffiltext{11}{NASA--Goddard Space Flight Center, Greenbelt, MD 20771}
\altaffiltext{12}{Observatories of the Carnegie Institution of
Washington, Pasadena, CA 91101-1292}
\altaffiltext{13}{Department of Astronomy, University of Virginia, 
Charlottesville, VA 22904-4325}
\altaffiltext{14}{Istituto di Astrofisica Spaziale e Fisica Cosmica, INAF, 
Via Gobetti 101, 40129 Bologna, Italy}
\altaffiltext{15}{National Optical Astronomy Observatories, Tucson, AZ
85726-6732}
\altaffiltext{16}{Department of Physics, University of Oxford, Oxford  
OX1 3PU, United Kingdom}
\altaffiltext{17}{NASA--Jet Propulsion Laboratory, Pasadena, CA 91109}
\altaffiltext{18}{Cerro Tololo Inter-American Observatory,
La Serena, Chile}
\altaffiltext{19}{School of Earth and Space Exploration, Arizona State
University, Tempe, AZ 85287-1404}
\altaffiltext{20}{NASA--Ames Research Center, Moffett Field, CA 94035}

\begin{abstract}
We investigate the ionization structure of the nebular gas in M83 using the line diagnostic diagram, \oiii/\hb vs. \sii/\ha    
with the newly available narrowband images from the Wide Field Camera 3 (WFC3) of the Hubble Space Telescope (HST). 
We produce the diagnostic diagram on a pixel-by-pixel ($0.2''\times0.2''$) basis and compare it with several photo- and shock-ionization models.
We select four regions from the center to the outer spiral arm and compare them in the diagnostic diagram. 
For the photo-ionized gas, we observe a gradual increase of the $\log($[O {\footnotesize III}]/H$\beta)$ ratios from the center to the spiral arm,  
consistent with the metallicity gradient,  
as the H {\footnotesize II} regions go from super solar abundance to roughly solar abundance from the center out. 
Using the diagnostic diagram, we separate the photo-ionized from the shock-ionized component of the gas. 
We find that the shock-ionized \ha emission ranges from $\sim$2\% to about 15--33\% of the total, depending on the separation criteria used. 
An interesting feature in the diagnostic diagram is an horizontal distribution around $\log($[O {\footnotesize III}]/H$\beta) \approx 0$. 
This feature is well fit by a shock-ionization model with 2.0 \Zsolar metallicity and shock velocities in the range of  250 km/s to 350 km/s. 
A low velocity shock component, $< $ 200 km/s, is also detected, and is spatially located at the boundary between the outer ring and the spiral arm. 
The low velocity shock component can be due to : 1) supernova remnants located nearby, 
2) dynamical interaction between the outer ring and the spiral arm, 3) abnormal line ratios from extreme local dust extinction.
The current  data do not enable us to distinguish among those three possible interpretations. 
Our main conclusion is that, even at the HST resolution, the shocked gas represents a small fraction of the total ionized gas emission 
at less than 33\% of the total. However, it accounts for virtually all of the mechanical energy produced by the central starburst in M83.


\end{abstract}


\section{Introduction}

The feedback from star formation activity into the interstellar medium (ISM) is a major, but still not fully characterized, mechanism that drives  
the formation and evolution of galaxies. In starforming sites, the stellar winds and supernovae explosions provide heat and momentum 
to the surrounding ISM changing its thermodynamic properties and kinematics, and sometimes driving galactic scale superwinds (Heckman et al. 1990, Martin et al. 2002). Such outflows from starforming regions carry metals that can enrich the intergalactic medium (IGM) and suppress further starforming activity (Oppenheimer \& Dav\'{e} 2006). Galactic scale outflows/winds have been called upon to account for the mass-metallicity relation, to shape the luminosity function of galaxies, especially at the faint end slope, and to account for the kinematics of neutral gas in Damped Lyman Alpha systems (Tremonti et al. 2004, Scannapieco et al. 2008, Hong et al. 2010). 

Despite the importance of understanding stellar feedback in the context of galaxy evolution, its observational constraints are still ill-defined. One of the largest unknowns is the ``energy efficiency" of feedback, i.e. the fraction of a starburst's mechanical energy that is available to drive large-scale outflows. This is linked, in the momentum-driven scenario, to the momentum transfer between the sources (supernovae, stellar winds) and the surrounding ISM coupled with radiative-driven pressure (Murray, Quataert, \& Thompson 2005).
On a more global scale, it is important to understand whether feedback and its efficiency can be linked to global galactic parameters like specific SFR or the depth of the galactic potential well (i.e. escape velocity). In this respect, local starburst galaxies are undoubtedly excellent laboratories to study the feedback interaction in great details. 

By investigating nearby galaxies, Martin (2005, 2006) suggests that a relation is  present between outflow speed and SFR, and that the kinetic energy contained in the outflowing material is a few percent of the available mechanical energy from massive star winds and supernovae. This implies that although the fraction of mechanical energy deposited in the ISM is high, only a small fraction is available to drive the outflow. 

Here we concentrate on the starburst and H {\footnotesize II} regions hosted in the center of the massive spiral galaxy M83, in order to investigate the interface between star formation and shocked gas not only in the powerful starburst, but also in the H {\footnotesize II} regions hosted in the spiral arms of the galaxy. We perform a differential analysis of the shocked gas surrounding the central starburst and the spiral arm H {\footnotesize II} regions, with the goals of deriving the fraction of energy deposited in shocks and their morphology.

The line diagnostic diagram from nebular line emission, 
\oiii/\hb vs. \sii   (or [N {\footnotesize II}](6583\AA))/\ha), 
has been used to study the starburst or AGN activity in galaxies (Baldwin, Phillips, \& Terlevich 1981, Kewley et al. 2001; hereafter K01) and to investigate the ionized gas structure of ISM for resolved regions (e.g. Calzetti et al. 2004; hereafter C04, Westmoquette et al. 2007). The HST's high angular resolution is crucial to study the ionized gas structure, especially for shock-ionized gas, since, in general, the shocked gas is distributed in a thin layer, easily confused with photo-ionization in  low resolution ground-based observations. We also need accurate ionization models since there is no strict boundary between photo-ionized gas and shock-ionized gas in the diagnostic diagram. In this paper we focus on the diagnostic diagram of M83 for the structure of ionized gas from the newly available data obtained with the Wide Field Camera 3 (WFC3) onboard the Hubble Space Telescope (HST). 

The galaxy M83 is the nearest face-on grand-design spiral with Hubble type SAB(s)c, at a distance of 4.47 Mpc (Kennicutt et al. 2008). There is an ongoing central starburst possibly fueled by gas inflow along the main stellar bar connecting the double circumnuclear rings found in the center (Trinchieri et al. 1985, Elmegreen 1998, C04). To study the non-photoionized gas (mostly shock-ionized gas driven by stellar feedback) in the central region, C04 analyzed Wide Field Planetary Camera 2 (WFPC2) images on a pixel-by-pixel basis and compared the line diagnostics with the photo-ionization model of K01 and shock-excited line ratios from Shull \& McKee (1979). To separate the shock-ionized gas from photo-ionized gas, C04 used the ``maximum starburst line'' of K01. Because the ``maximum starburst line" is a conservative boundary for shocks, the shock measurement from those authors is a lower limit to the actual fraction of shock emission. They find that the fraction of shock-ionized \ha luminosity to photo-ionized luminosity is $\sim 3$\% in M83. Unlike the dwarf galaxies, also analyzed in that paper, the geometrical morphology of shock ionized gas in M83 was compact, rather than diffuse and shell-like, possibly due to the deeper potential well in this more massive galaxy. 

We provide, in this paper, the observed line diagnostics of the center and the inner spiral arm of M83, 
and compare the data with theoretical expectations for the line ratios. 
In section \S 2, we describe the data set and the reduction procedures, 
including a detailed discussion of how to remove emission line contamination from broad/medium band filters and [N {\footnotesize II}] contamination from the \ha line filter.
Then we present our results on the ionized gas structures and interpretations from theoretical models in section \S3. The summary follows in section \S 4.

\section{Observations and Data Reduction}

\subsection{The WFC3 data set}

Our WFC3 data are part of the WFC3 Science Oversight Committee (SOC) Early Release Science (ERS) program 
(program ID11360, PI: Robert O'Connell). The observed field, centered at R.A. = 13:37:04.42, 
DEC. = -29:51:28.0 (J2000), covers the nuclear starburst region and the inner part of northeast spiral arm of M83. 
Three or four (depending on the filter) spatially dithered exposures were used to remove cosmic rays 
through the MultiDrizzle software (Fruchter et al. 2009). 
We adopt the PHOTFLAM values, available in URL:http://www.stsci.edu/hst/wfc3/phot\_zp\_lbn,  to convert instrumental data number(DN) to physical flux. We use the flux calibration value of F658N for F657N, for which a calibration is not available, since their values 
for line emissions are similar within a few percent. The detailed information about our data set is summarized in Table 1. 
Before we derive line ratio maps, we rebin the images by $5\times5$ pixels to reduce registration errors. The binned pixel size is $0.2''$ or $4.3$ pc in physical scale; i.e. smaller than the typical size of an H {\footnotesize II} region, but comparable to the physical size of star clusters (Maiz-Apellaniz 2001).

\subsection{Continuum Subtraction and [N {\footnotesize II}] Correction}

The F487N and F502N narrowband filters each contain one line emission per filter 
(\hb and \oiii$~$, respectively,  redshifted to the recession velocity of M83, 513 km/s). 
F657N and F673N contain multiple lines. In the case of F673N, the lines are both [S {\footnotesize II}], 
from the 6716\AA$~$ and 6731\AA$~$ doublet, so we will simply consider the sum of the two. 
In the F657N filter, the [N {\footnotesize II}] doublet at rest frame 6548\AA$~$ and 6584\AA$~$ 
is included in the filter bandpass in addition to \ha. 
Hence, we have simpler filter throughput equations to obtain \hb and \oiii$~$ fluxes than \ha and \sii. 
As a trade-off, however, the stellar continuum contained in the F487N and F502N filter is more affected by variations in the stellar population mix and the local dust 
extinction than the redder lines, 
which means their stellar continuum estimates have worse intrinsic accuracy than F657N and F673N. Furthermore, the F555W filter, 
which we use as the broad-band continuum filter to estimate the stellar contribution to F487N and F502N, is self-contaminated 
by \hb and \oiii; we assume [O {\footnotesize III}](4959\AA) is negligible compared to the two other lines. 
Therefore, we have to correct our filter fluxes for the following: (1) contamination of \hb and \oiii$~$ in the F555W filter, 
(2) [N {\footnotesize II}] contamination in F657N. 

To subtract the \hb and \oiii \msp lines from the F555W image, we apply an iterative method which can be schematically described as: 
\begin{eqnarray}
Cont_{0} &\leftarrow&  F555W  \nonumber \\
H\beta_{i} &\leftarrow&  F487N -  Cont_{i} \nonumber \\
\big[O ~ III\big]_{i} &\leftarrow& F502N -  Cont_{i} \nonumber \\
Cont_{i+1} &\leftarrow&  F555W - H\beta_{i} - [O ~ III]_{i} \nonumber
\end{eqnarray}
where each term represents the flux calibrated image for the denoted line emission and filter. 
In this method, the starting value for the stellar continuum is the uncorrected F555W, $Cont_0$. 
With it, we obtain the first guesses to the \hb and \oiii$~$ emission line images by 
subtracting the stellar continuum from the respective narrow-band filters. 
Then, by subtracting those first-guess line emissions from f555w, we can derive the first order corrected continuum, $Cont_1$. 
We can iterate the procedure until the image converges. This procedure is similar in concept to the iterative procedure applied in Mackenty et al. (2000) for NGC4214.

To quantify the improvement of the iterative method, we compare the corrected F555W continuum 
after each iteration with the F547M image available in the HST archive 
for the central region of M83, since F547M is a line emission-free filter. 
We produce a residual image, $\frac{ F547M - \mu F555W }{F547M}$, for each iteration step to investigate the 
effect of each iteration on the F555W image. The parameter, $\mu$, is chosen to minimize the difference between the images in the two filters and is related to the scaling factor for the stellar continuum subtraction from narrow-band images. 

\begin{figure}[t]
\centering
\includegraphics[height=4.0 in]{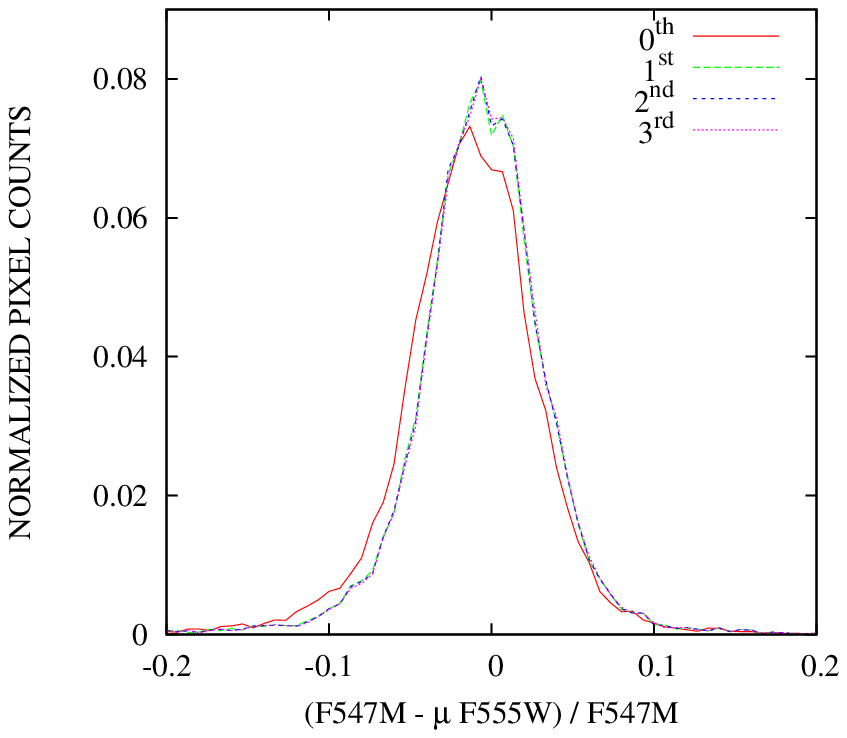}
\caption{The pixel histograms of the residual images, $\frac{ F547M - \mu F555W }{F547M}$, 
for each iteration of the emission line subtraction algorithm from F555W. 
The parameter, $\mu$, is chosen to 
minimize the difference between the images in F547M and F555W 
using the same method used for the continuum subtraction in the narrow band images. 
The histograms show the overall difference between the images in the F547M filter and 
in the corrected F555W filter at each iteration. The original, $0^{th}$, histogram 
is skewed towards a negative mean value, where higher order iterations produce 
more symmetric, and similar to each other, histograms. 
The overall excess of negative pixels in the $0^{th}$ image is the amount of line emission contamination in F555W. 
The improvement of the $1^{st}$ corrected image from the original one is large, 
and further iterations do not improve on that by an appreciable amount . 
}\label{fig:iterhisto}
\end{figure}

\begin{figure}[t]
\centering
\includegraphics[height=4.0 in]{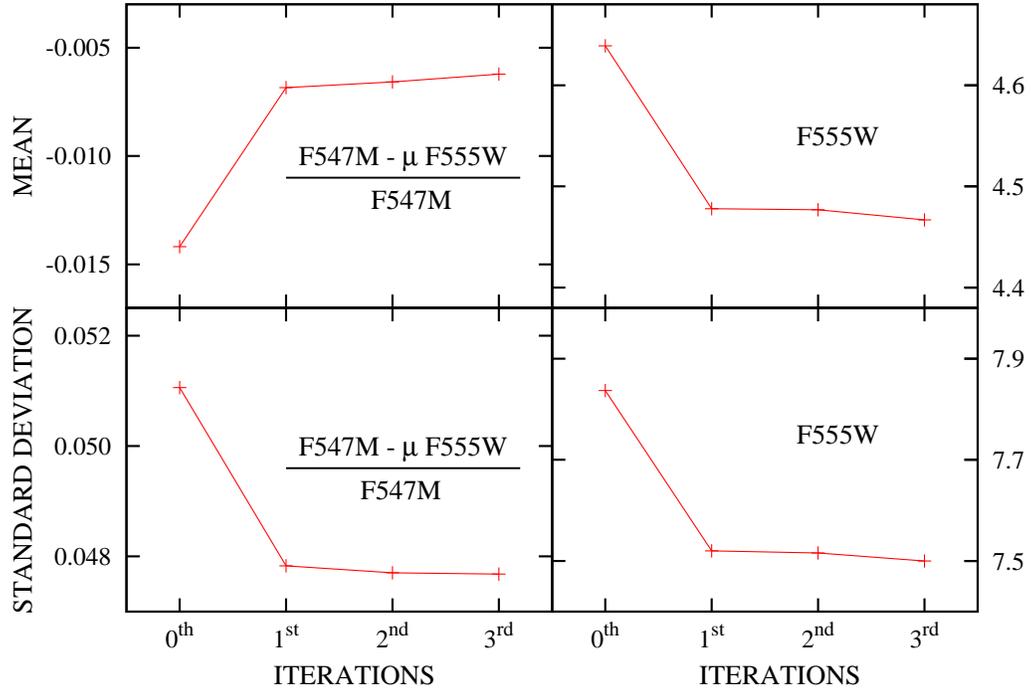}
\caption{ The mean (top) and the standard deviation (bottom) of the residual images (left) , $\frac{ F547M - \mu F555W }{F547M}$,  
and the corrected F555W images (right) as a function of the number of iterations. As 
shown in Figure ~\ref{fig:iterhisto}, the $1^{st}$ corrected image shows a measurable improvement, 
while further iterations do not change the image quality in a measurable fashion. So the second iteration is enough 
for the final derivation of the \hb and \oiii \msp line emission images. 
}\label{fig:itertrend}
\end{figure}

Figure~\ref{fig:iterhisto} shows the pixel histogram of residual image for each iteration. 
The uncorrected, $0^{th}$ iterated, residual image shows a negative mean which indicates excess flux from \hb and \oiii \msp in F555W. 
After the $1^{st}$ iteration, the residual images show a more symmetric pixel histogram than the uncorrected one 
showing that the iterations remove the excess of line emission in F555W. 
The difference between the uncorrected and the 1st corrected is appreciable especially in localized regions, where flux changes up to 100\% have been measured. Further iterations do not change the F555W image appreciably. To quantify the difference for 
each iteration step, we shows the mean (top) and the standard deviation (bottom) of the residual images (left) and the corrected F555W (right) in Figure~\ref{fig:itertrend}. We can observe that 1, at most 2, iterations are sufficient to reach convergence.  
The converged residual histogram in Figure~\ref{fig:iterhisto} can be considered as an intrinsic difference between the two filters. 
We take as final \hb and \oiii$~$ emission line images those resulting from the second-iterated F555W stellar continuum image.

The optimal scale factor to be applied to the broad band images for continuum subtraction in the narrow band filters is found with the skewness transition method (Hong et al. 2011; in prep). The skewness is defined as  
\begin{equation}
skewness = \frac{1}{N-1} \sum_{i=1}^{N} \big( \frac{x_i - m}{\sigma}  \big)^3 \nonumber
\end{equation}
, where $m$ is the mean, $\sigma$ is the standard deviation, and $N$ is the number of pixels in a sample image. 
The skewness is generally used to measure asymmetry of a statistical distribution. 
For symmetric functions, the value of the skewness is zero. If a distribution has a long ``right'' tail, the skewness of the distribution 
is positive; i.e. positive-skewed. For a distribution with a long ``left'' tail, the skewness is negative; i.e. negative-skewed. 
For an image of blank sky, the pixel histogram is generally gaussian (or poissonian). So the skewness is zero 
(or a small positive number). Additional astronomical sources in a blank field, therefore, are weighing on the right-hand side of 
pixel distribution which makes the distribution ``positive-skewed''. Since the continuum subtraction process removes stellar flux 
from the narrow-band image, increasing the subtraction weight decreases the skewness value.   
In a perfectly-subtracted narrow-band image, the skewness of the image pixels that do not contain line emission is zero for a gaussian background. Our method attempts to recover the optimal stellar continuum subtraction value by exploiting this property of the skewness. 
Our simulations (Hong et al. 2011) show that the skewness function shows a ``transition'' in correspondence of the optimal scaling factor for the stellar continuum. So, by locating where this transition occurs in our HST images, we have derived the optimal scaling factor for each narrow-band filter. 

It should be noted that the use of a single scaling factor across each image assumes that large color gradients (due to either changes in the stellar population mix or in the dust extinction) are not present across the broad band filters. For F657N and F673N, this assumption is mitigated by interpolating the two broad band filters, F814W and the corrected F555W, to derive stellar continuum images at the appropriate wavelength.  Furthermore, except for a few regions in the nucleus of M83, colors do not change dramatically across the F555W or F814W filter band passes. 
We use a single continuum image interpolated at the intermediate wavelength 660 nm for both of F657N and F673N. Since the widths of the broad-band images are larger than 100 nm, the interpolated image at 660nm can be used as continuum for nearby emission lines located within a couple of 10 nm.  

We produce two subtracted versions (optimized subtraction for the central region and the spiral arm each) for \oiii$~$ and \hb, since the bluer continuum F555W suffers from more local deviation from color differences of background stellar population than the redder filters. We take a single optimized subtraction for \ha and \sii. We consider potential changes in color within the broad band filters as a source of error in our photometry. 

We assume that [S {\footnotesize II}](6716/6731) $\approx 1.2$, since the ratios are 
between 1.0 and 1.4 in the ISM of M83 (Bresolin \& Kennicutt 2002; hereafter BK02). 
For the [N {\footnotesize II}] correction to \ha, we adopt the spectroscopic flux ratios, 
[N {\footnotesize II}]/\ha = 0.42 for the spiral arm and 0.54 for the center from BK02. 
We, thus, generate two ``H$\alpha$'' only maps; one with the [N {\footnotesize II}] subtraction 
optimized for the central region, and one optimized for the spiral areas 
(Note that \hb and \oiii \msp have two optimized versions due to different continuum subtraction, while for \ha due to different [N {\footnotesize II}] correction). 
The [N {\footnotesize II}]/\ha ratios range from 0.40 to 0.47 in the spiral arm and 
from 0.53 to 0. 56 in the center. 
The difference of the corrected \ha fluxes within each of these ranges is less than 10 \%, which is small 
enough to be negligible in our analysis. C04 used the relation [N {\footnotesize II}] $\approx 2 \times$[S {\footnotesize II}] 
too as an alternative method for [N {\footnotesize II}] correction . The ratio [N {\footnotesize II}]/[S {\footnotesize II}] 
has been observed to be relatively constant, since the ratio has less dependences on abundance 
and extreme variations in UV radiation (Rand 1998, Kewley \& Dopita 2002). 
C04 discusses the [N {\footnotesize II}] correction in more details. Our adopted method produces results similar to those of C04. 
A summary is given in Table 1 of the images or combination of images used for the stellar continuum subtraction 
of each narrow-band image.

\section{Results}

\begin{figure}[t]
\centering
\includegraphics[height=5.5 in]{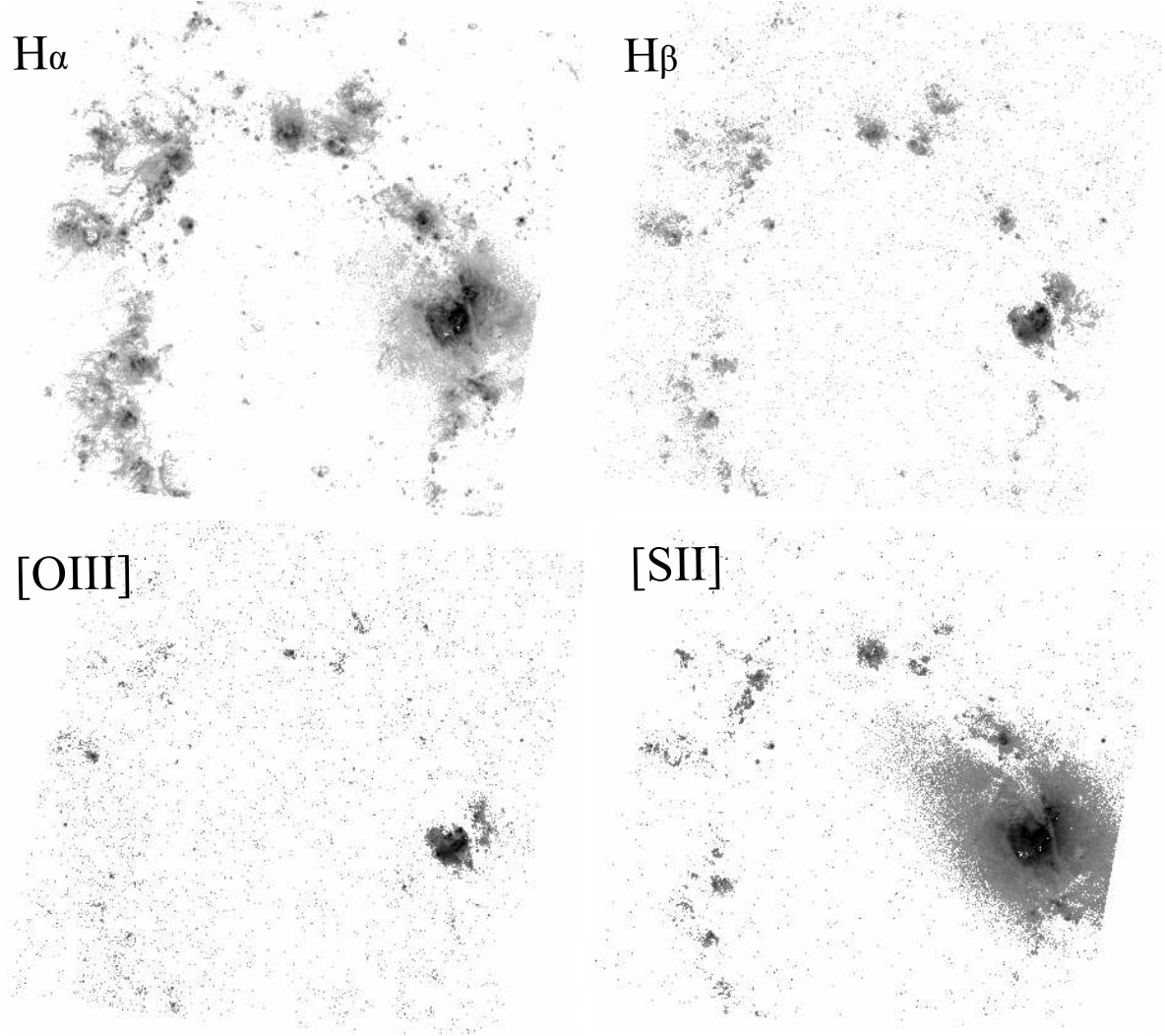}
\caption{The 4 $\sigma$ images for \ha , \hb , \oiii, and \sii . The images of \hb and \oiii $~$ shown here are the versions
 where the stellar-continuum subtraction has been optimized for the nuclear region. The field shown here is the entire field of view of WFC3/UVIS. 
}\label{fig:m83allimages}
\end{figure}

Figure~\ref{fig:m83allimages} shows the continuum subtracted images for \ha, \hb, \oiii, and \sii \msp with 4$\sigma$  cuts 
(see Table 1 for a list of the 1 $\sigma$ flux levels). 
For \hb and \oiii, the stellar continuum subtractions optimized for the central regions are presented. 
We take $4\sigma$ cuts for all the narrowband images because of the weakness of \oiii$~$ emission. 
Deeper rejection cuts would remove enough of the [O {\footnotesize III}] emissions to remove most of the shock-ionized signal.  
As shown in Figure~\ref{fig:m83allimages}, the surface brightness of \oiii \msp is 
the major limiting factor to the number of data points in the diagnostic diagram. To compensate for this limitation, we also utilize the shock criteria from the \sii/\ha ratio. This method is less rigorous for shock separation than the diagnostic diagram but it can include more area for our investigation, since \sii \msp is substantially 
brighter than \oiii \msp in M83. 

\begin{figure}[t]
\centering
\includegraphics[height=5.7 in]{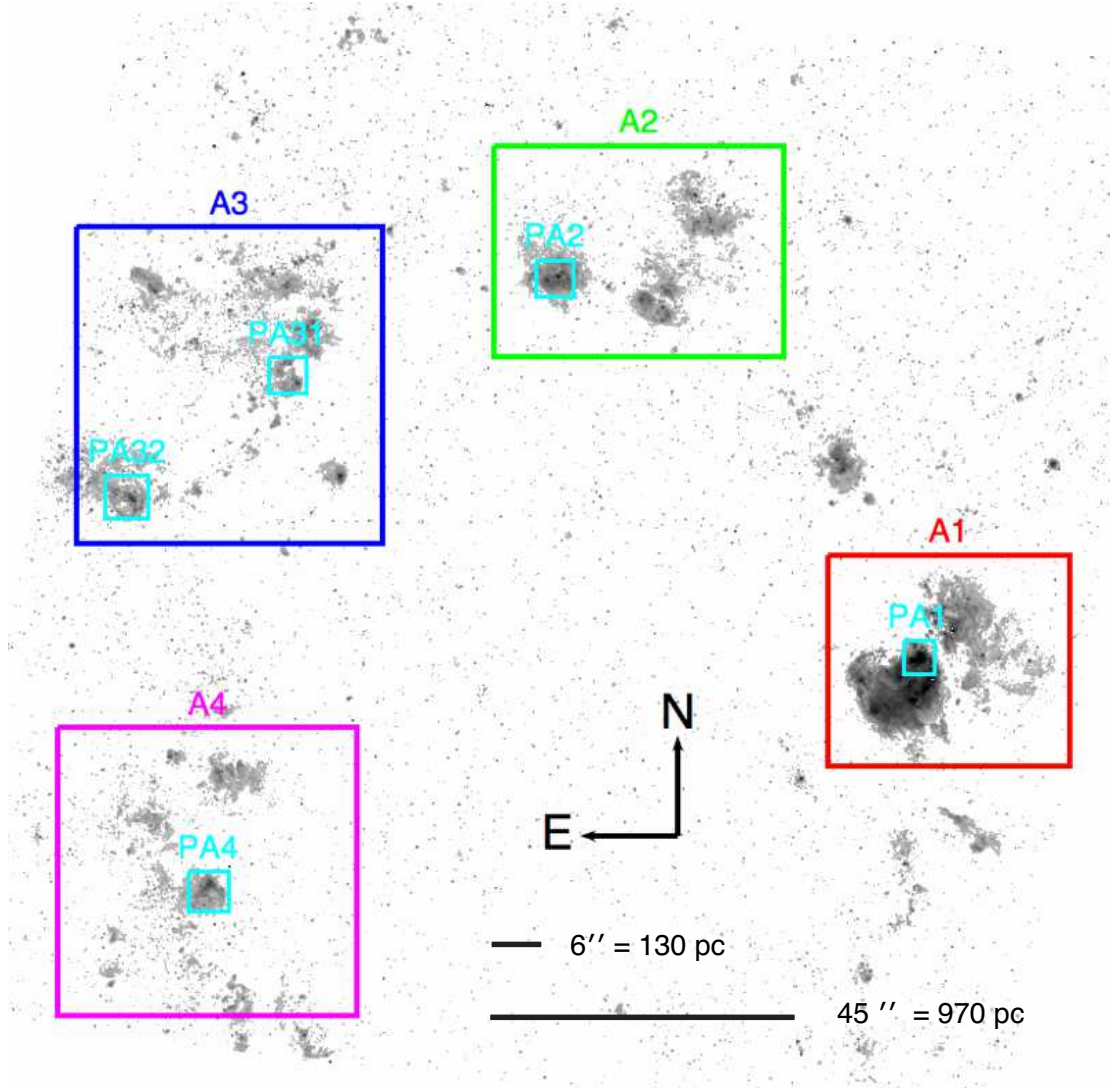}
\caption{The \hb image (4$\sigma$) and the locations of regions, A1, A2, A3, and A4. The smaller boxes, PA1,.., and PA4 are selected from the ``P'' box in Figure~\ref{fig:diag}. 
PA1 and PA4 correspond to the regions ``A'' and ``9'', in BK02, respectively. 
}\label{fig:m83area}
\end{figure}

We select four regions within the field of view of our narrow 
band images, which are plotted as squares overlaid on the H$\beta$ 
image in Figure~\ref{fig:m83area}. The four sub-regions are 
labeled A1(red), A2(green), A3(blue), and A4(magenta). These regions are bright enough in \oiii$~$ to enable analysis in the diagnostic diagram. They are  geometrically separated from each other 
and span a total galactocentric separation of 2.5 kpc, 
which can probe the change of physical properties from the center to the spiral arms.


In star forming regions, ultraviolet(UV) photons from stellar objects and hot coronal gas heated by shocks are the major source of 
ionization of the surrounding ISM. The total luminosity and hardness of the UV radiation are determined by a combination of 
 star forming history, metal abundance, stellar atmosphere model, and stellar evolution tracks. We will refer to this as ``photo-ionization''. 
The other mechanism is collisional ionization that takes place in shocks caused by stellar winds and supernovae. 
The stellar feedback energy from supernovae and stellar winds heats up the 
surrounding gas and produces hot bubbles in star forming sites. Since the temperature in hot bubbles is over a million degrees, the bubble 
is generally faint in the optical, while it is bright in the X-ray. The over-pressure set up in the bubble by the feedback energy pushes 
out the surrounding gas and produces shocks in the interface between the bubble and the ambient ISM. When the shock becomes 
radiative in the early adiabatic phase, the upstreaming radiation from the shock layers ionizes the pre-shock gas. Unlike stellar emission, the radiation from shock layers 
is dominated by free-free emission of hot coronal gas, so the properties of the shock emission are different from those of the stellar emission. We term this 
as ``shock-ionization''. The diagnostic diagram, \oiii/\hb vs. \sii/\ha , has been traditionally used to discern these two ionization mechanisms and many ionization models provide predictions based on this diagnostic diagram for comparison with observed data.  

\begin{figure}[t]
\centering
\includegraphics[height=4.2 in]{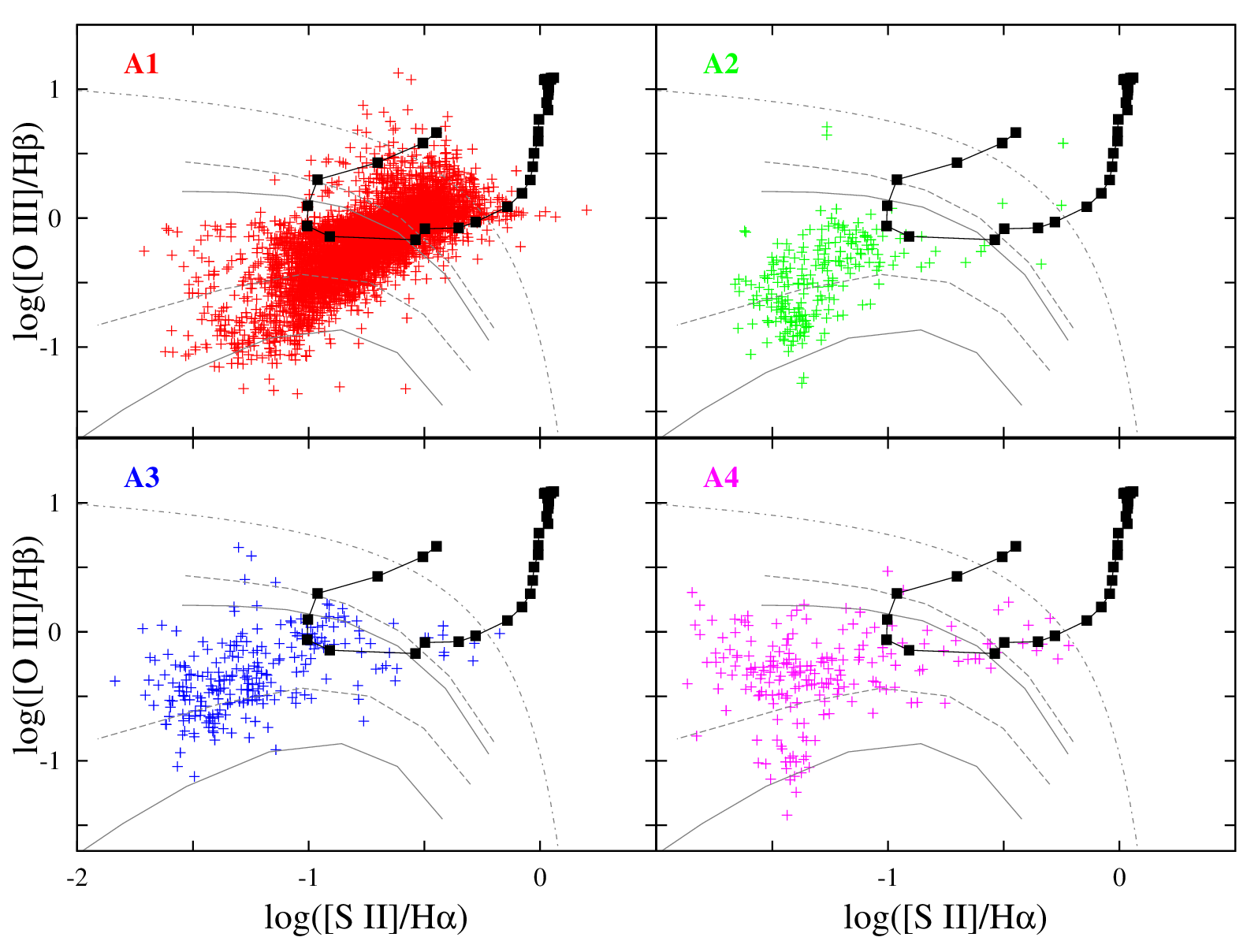}
\caption{The line diagnostic diagram, \oiii/\hb vs. \sii/\ha, for each area 
delineated by a large box in Figure~\ref{fig:m83area}, A1(top-left), A2(top-right), A3(bottom-left), and A4(bottom-right).  
The data are color-coded according to the colors of the box they belong to Figure~\ref{fig:m83area}. 
Sample theoretical models for photo-ionized gas (grey lines, K01) and 
shock-ionized gas (black squares joined by lines, A08) are also shown 
in each panel. A detailed description of the models is given in the text and in 
Figure~\ref{fig:diag}.
}\label{fig:diag4regions}
\end{figure}

Figure~\ref{fig:diag4regions} shows the distribution of the data points ($0.2''$ pixels) from 
each of the four subregions on the diagnostic diagram, together with 
theoretical tracks from various ionization models of K01. Figure~\ref{fig:diag} 
shows the data points from all four subregions on a single diagram. 
By comparing the data points with the theoretical grids, we can infer many properties for the ionized gas in M83.  
In the following sections we present in detail results about the photo-ionized gas, the shock-ionized gas, and 
the implications of different choices for separating the two components.

\subsection{Photo-ionized Gas}

\begin{figure}[t]
\centering
\includegraphics[height=4.2 in]{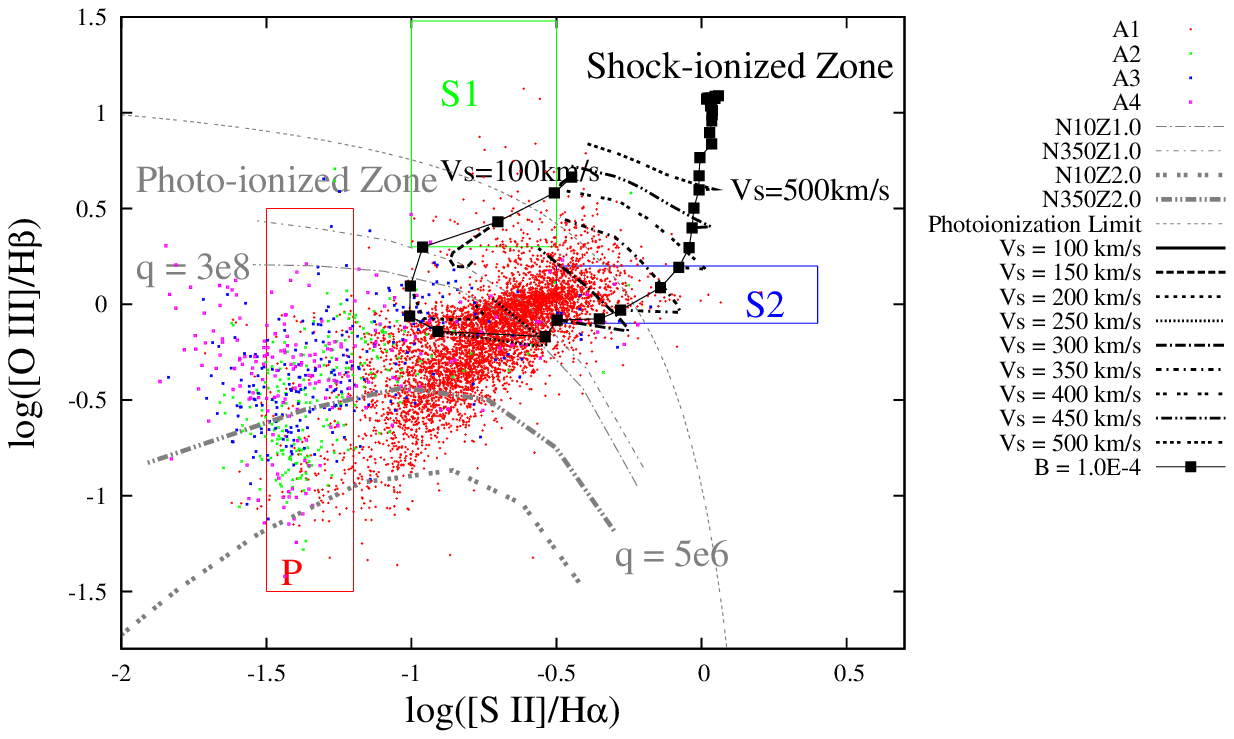}
\caption{The line diagnostic diagram, \oiii/\hb vs. \sii/\ha, for each area of Figure~\ref{fig:m83area}, A1(red), A2(green), A3(blue), and A4(magenta). 
The grey lines represent photo-ionization models of constant star formation from K01. 
The ``Photoionization Limit'' listed in the legend to the right of the figure is the conservative limit for extreme photo-ionization introduced in K01. 
The model name N10Z1.0 represents the case of solar metallicity and an ISM density of 10 cm$^{-3}$, 
and is plotted for the range 5E+6 to 3E+8 
for the dimensionless ionization parameter q. 
Others follow the same naming convention and are plotted for the same range of ionization parameters (grey lines).   
The black lines represent the shock-ionization model with twice solar metal abundance from A08 (Shock + Precursor model). 
The ``B=1.0E-4'' line shows the line ratios for various shock velocities 
(100, 125, 150, ... , 900km/s at 25 km/s intervals; black squares) with the $10^{-4} \mu$G magnetic field. 
The side branches from the ``B=1.0E-4" line of selected shock velocities (100,150, ... , 500 km/s; side branches from black squares) 
show the effect of increasing magnetic fields from $10^{-4}$ to 10 $\mu$G. 
The boxes, ``P'',``S1'' and ''S2'', are described in the sections, 3.1 (P) and 3.3 (S1 and S2) respectivily. 
}\label{fig:diag}
\end{figure}

Figure~\ref{fig:diag} shows the overall distribution of the ionized gas in M83 on the diagnostic diagram. 
The grey lines are photo-ionization models from K01 and the black lines are 
shock-ionization models from Allen et al. (2008; hereafter, A08), which will be described later in the section. 
The photo-ionized grids adopted from K01 are based on the stellar population synthesis model STARBURST99 
and the gas ionization code MAPPINGS III (Leitherer et al 1999, Binette et al. 1985, Sutherland \& Dopita 1993). 
The spectral energy distributions (SED) generated from STARBURST99 provide the photo-ionization source and 
the MAPPINGS III code calculates the ionization states and the line emission fluxes. 
From those, we can derive the theoretical line ratios for photo-ionization. 
The plotted photo-ionization grids in Figure~\ref{fig:diag} are calculated assuming a constant star formation history, 
Geneva stellar evolution tracks (Schaller et al. 1992), and Lejeune stellar atmosphere models (Lejeune et al. 1997); see K01 for more details. 
While the assumption of constant star formation may not strictly apply to the spiral area H {\footnotesize II} regions, 
it is a good representation of star formation in the starburst nucleus of M83 (C04). The case of instantaneous burst models 
will be discussed later in this section. 

The selected ISM densities are 10 and 350 $cm^{-3}$ which we take as representative of the ISM conditions in M83, 
and the metallicities are chosen to be 1.0$Z_{\odot}$ and 2.0$Z_{\odot}$, 
the range observed in the galaxy (BK02). The dimensionless ionization parameter, $q$, defined as the ratio of mean photon density to mean atom density,  ranges from $5.0\times10^{6}$ to $3.0\times10^{8}$ for each model. 
The line called ``Photoionization Limit"  in the legend of Figure~\ref{fig:diag} 
is the conservative track termed ``Maximum Starburst Line" 
in K01. Above and to the right of this track, line ratios can not be explained by photo-ionization. 
Since the outputs of STARBURST99 depend on a variety of inputs, such as stellar evolution tracks, 
stellar atmosphere models, and star formation history, 
we can use our theoretical grids primarily for qualitative comparisons rather than a quantitative analysis. 

One of the interesting results from the photo-ionized zone in Figure~\ref{fig:diag} is that the observed [O {\footnotesize III}]/\hb$~$ ratios at low [S {\footnotesize II}]/\ha values span the range covered 
by the photo-ionization models for different values of the metallicity; higher \oiii/\hb \msp ratios correspond to lower metallicity in the models.  
Generally \oiii/\hb$~$ ratios are not enough to determine the metallicity of a region (Kewley \& Dopita 2002), 
except in the range between solar abundance to super-solar abundance. 
The red ``P'' box in Figure~\ref{fig:diag}  is a photo-ionized region, 
where the \oiii/\hb$~$ line ratio is sensitive to metallicity. We re-project the pixels of the P box 
on the line emission images and select small subregions, labeled as PA1, PA2, PA31, PA32, and PA4, as shown in Figure~\ref{fig:m83area}, to investigate the relation between the observed \oiii/\hb ratio and metallicity. 
Those regions are bright in each emission line and geometrically clustered in compact regions. 
The sizes of the regions are close to the size of the spectroscopic aperture used in BK02. In addition, PA1 and PA4 correspond to the regions, 
``A'' and ``9'' in BK02, respectively. 

In the P box, we can also observe a weak rank-ordering of the  [O {\footnotesize III}]/\hb ratios according to the regions they belong to : $A1 > A2 > A3 \approx A4$. 
Interestingly, that is the exact order of distance from the center. This relation between the [O {\footnotesize III}]/\hb ratios and the distance from the center 
appears consistent with the metallicity gradient from the center to the spiral arms in M83. 

\begin{figure}[t]
\centering
\includegraphics[height=4.2 in]{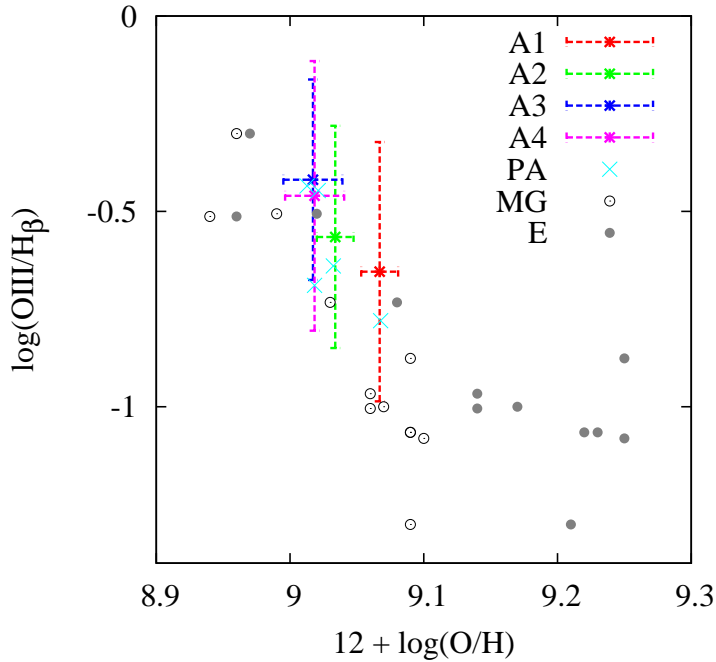}
\caption{ The ratio of [O {\footnotesize III}]/H${_\beta}$ vs. the oxygen abundance. The black and grey points are 
reproduced from Table 3 in BK02. The name ``MG'' and ``E'' indicate the different oxygen abundance calibrations from McGaugh (1991) and Edmunds (1989).  The smaller regions, PAs, have the angular sizes around $5''$ which is somewhat larger than, but 
still comparable to, the spectroscopic slit size $2''$ in BK02. 
The mean ratios for A1, A2, A3, and A4, are slightly larger than the spectroscopic results.
For the subregions, PA1,..., and PA4, the mean values are consistent with 
BK02. The vertical errorbars of the PA regions, the standard deviations of the line ratio distributions for PA regions, are as large as their parent regions, A1, ... , and A4, so we omit the vertical error bars for the PA regions. 
Since the PA regions are chosen for their brightness and compactness relative to 
A1,..., and A4, the higher [O {\footnotesize III}]/\hb ratios in A1,..., and A4, can be 
due to  gas that is too faint and diffuse for spectroscopic study. 
}\label{fig:oiiivsmetal}
\end{figure}

To verify this, we present the observed relation between oxygen abundance and [O {\footnotesize III}]/\hb ratios in Figure~\ref{fig:oiiivsmetal}. 
The black (named ``MG'') and the grey (named ``E'') points  are borrowed from the spectroscopic results of BK02. The names ``MG'' and ``E'' indicate the different oxygen abundance calibrations from McGaugh (1991) and Edmunds (1989). 
The figure shows the direct correlation between [O {\footnotesize III}]/\hb ratio and oxygen abundance. 
To compare our data with those from BK02, we convert the distance of each region to the oxygen abundance using the relation between distance and oxygen abundance in BK02 and overplot them in Figure~\ref{fig:oiiivsmetal}. The horizontal error bar is due to the physical size of each region and the vertical error bar represents the standard deviation of the line ratio distribution among the pixels within each region.  
Since our resolution is much higher than the spectroscopic aperture, $0.2''$ for each pixel, the line ratios are distributed in a broad range 
for the A1, ..., A4 regions, as well as the PA1, ..., PA4 regions. The overall trend is consistent with the same metallicity trend as BK02. So the metallicity gradient from center to spiral arm appears to cause 
the spread in [O {\footnotesize III}]/\hb values in the ``photo-ionized'' area of the diagnostic diagram of Figure~\ref{fig:diag} at low [S {\footnotesize II}]/\ha values. 

\begin{figure}[t]
\centering
\includegraphics[height=6.0 in]{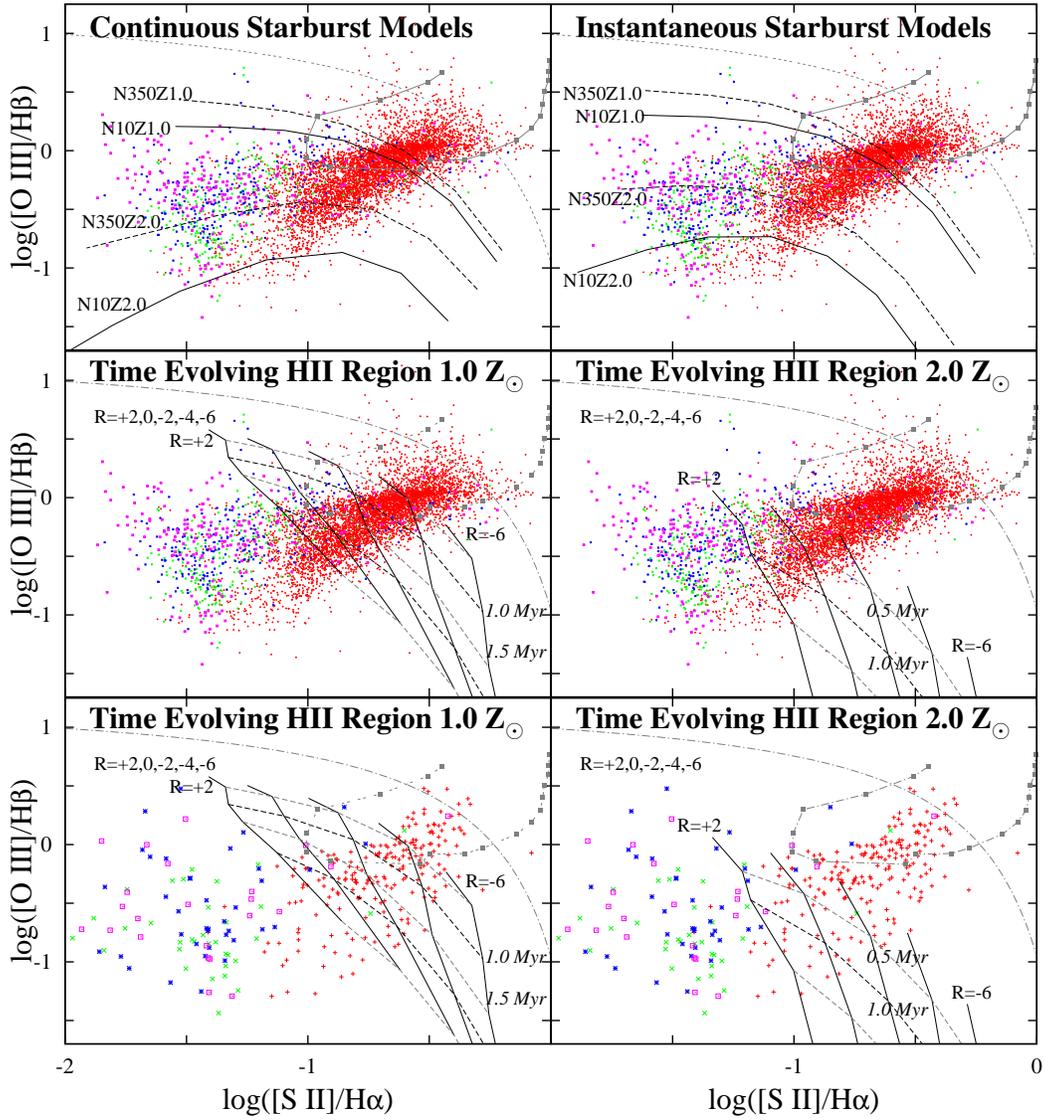}
\caption{ Top: The continuous starburst models (left) and the instantaneous starburst models (right) of K01. Middle: The time evolution tracks of H {\footnotesize II} regions from Dopita et al. (2006); solar abundance (left) and twice solar abundance (right). Bottom: The rebinned data points by $7\times7$ pixel square ($35\times35$ bin for the original WFC3 pixels)  corresponding to $1.4''$. The rebinned resolution is the spatial resolution of Rich et al. (2010) and a generally achievable one in ground-based observation. 
}\label{fig:photozoom}
\end{figure}

To investigate in detail the physical properties of the photo-ionized gas in M83, we present additional photo-ionization models in Figure~\ref{fig:photozoom}. The two top panels show two different star formation history models; constant star formation (left) and instantaneous burst (right) from K01. 
The constant star formation model assumes durations larger than 8 Myr, when the dynamical balance between the stellar births and deaths is set up for UV radiation. The instantaneous burst model is the case for zero age stellar population, which is the extreme end in star formation history. Thus, these two models bracket a range of star formation histories for our regions. 
Both show similar metallicity dependence; lower [O {\footnotesize III}]/\hb ratio for higher metallicity. 
The constant star formation model with  2.0 $Z_{\odot}$ marks the lower envelope to the photo-ionized data, thus  fully including all pixels in A1. 

Dopita et al. (2006; hereafter DP06) introduced models for time-evolving H {\footnotesize II} region. 
While K01 treats the ionization parameter as an independent parameter, DP06 calculates 
the ionization parameter as a function of the age of the stellar cluster. 
The DP06 models are based on a one dimensional spherical geometry for the evolving H {\footnotesize II} region. The radius and 
pressure evolution are derived from the works of Castor et al. (1975) and Oey \& Clarke (1997, 1998): 
\begin{eqnarray}
R &=& \big(\frac{250}{308\pi}\big)^{1/5} L_{mech}^{1/5} \rho_{0}^{-1/5}t^{3/5} \nonumber \\
P &=& \frac{7}{(3850\pi)^{2/5}} L_{mech}^{2/5} \rho_{0}^{3/5}t^{-4/5} \nonumber
\end{eqnarray}
where $L_{mech}$ is the instantaneous mechanical luminosity from the cluster, $\rho_{0}$ is the ambient ISM density, and $t$ is the time. 
DP06 found that the ionization parameter in their models depends on the mass of the stellar cluster and the pressure of the surrounding ISM, 
and it scales as $q \propto M_{cl}^{1/5} P_{0}^{-1/5}$, where $M_{cl}$ is the initial mass of the stellar cluster and $P_{0}$ is the initial ISM pressure surrounding the stellar cluster. Hence, the ratio, R$ = \log_{10}[ (M_{cl}/M_{\odot})/(P_{0}/k) ]$ where $P_{0}/k$ is measured in cgs units ($cm^{-3}$ K), uniquely determines the time-evolving track of the ionization parameter as an initial condition (see DP06 for more details). 
The middle panels in Figure~\ref{fig:photozoom} show the DP06 models for two 
different metallicities, 1.0 $Z_{\odot}$(left) and 2.0 $Z_{\odot}$(right). 
We also compare these models with re-binned data (see bottom panels of Figure~\ref{fig:photozoom}) 
to investigate the impact of the binning size on our results. 
In all cases analyzed, we find that an evolving HII region scenario at constant 
metallicity does not fully reproduce the observed trend in the data, although each covers a 
portion of the locus occupied by the data points. The full range of the data can be 
covered by varying the metal abundance by more than a factor 10, which we consider unlikely 
given the relative small size of our regions ($\approx$970~pc). At this stage, the best 
model to fit the photo--ionized gas appears to be the continuous starburst scenario.

The instantaneous models assume zero-aged stellar population and DP06 models track the age effect on such instantaneous models using the evolution dynamics described by the above equations. These two models are hard to reconcile with observations 
if the stellar population contains a mix of stellar clusters with various ages or if the ionization parameter 
can not be derived from the equations above. Those equations are exact only in the case of an isolated H {\footnotesize II} region. 
For a starburst region, the photo-ionized gas is diffuse and distributed on a large scale plus the region itself 
is the complex superposition of many H {\footnotesize II} regions.
For this case, the ionization parameter is better treated as an independent parameter rather than derived from an isotropically expanding sphere. 
The age distribution of stellar clusters in M83 approximately follows a power-law distribution (Chandar et al. 2010): 
there are many clusters formed within $10^7$ yrs, and the power-law extends by a few million years. 
Therefore, the combination of multiple stellar population ages and complex geometry of the photo-ionized gas can be the reasons why the continuous models fit the observation more closely.

\subsection{Transition from Photo-ionization to Shock-ionization}


The adopted separation criteria for the shock and photo-ionized gas 
need to be discussed in this kind of study, because there is no objective separation between the two phases in the diagnostic diagram. 
The maximum starburst line, or ``Photoionization Limit'' in Figure~\ref{fig:diag}, is a conservative limit (see K01 and Kauffmann et al. 2003), 
above and to the right of which 
line ratios can not be produced by photo-ionization. This provides a lower limit to the amount of shock-ionized gas; in Table 2, we list, for this discriminant (called M1), the \ha luminosity, fraction of the total luminosity, and fraction of the total area occupied by shocks. 
A second discriminant can be provided by photo-ionization models matched to the solar-to-super-solar 
metallicity of M83. If we choose, conservatively, the N350Z1.0 
model of Figure~\ref{fig:diag}, the fraction of \ha luminosity and area occupied by shocks increases dramatically (M2 in Table 2). 
In this case, even the H {\footnotesize II} regions located along the spiral arm outside the central region show evidence for a shock component, albeit at the level of $\approx 1$\%. 
Conversely, in the central starburst, about 15\% of the total \ha luminosity is due to shocks.  
A third discriminant (called M3) can be provided by shock-ionization models. Data points will be 
considered shocks if they lie within the region covered by shock+precursor models. The M3 criterion is 
the most generous, among the three discussed so far, in including pixels from each region among the 
`shocks', it will provide a robust upper limit to our shock estimates. The shock H$\alpha$ luminosity 
in the spiral arm regions is now 4\% (A3) or less of the total \ha luminosity, still is not a significant 
amount. A much larger fraction, about 33\% of the total, is derived for the central starburst, the most 
generous estimate among all criteria.

\begin{figure}[t]
\centering
\includegraphics[height=4.0 in]{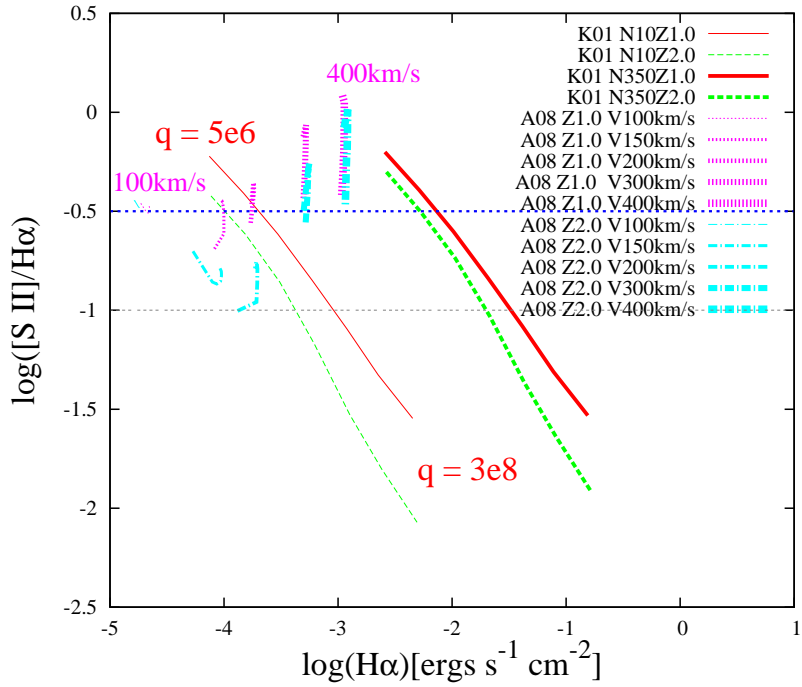}
\caption{The $\log$([S {\footnotesize II}]/H$\alpha$) vs. the normalized \ha flux from K01 and A08 models. 
From the $\log$ ([S {\footnotesize II}]/H$\alpha$) ratios, we can define three zones; 
1) shock-ionized zone: $\log($[S {\footnotesize II}]/H$\alpha) > -0.5$; 
2) mixing zone:  $-1 < \log($[S {\footnotesize II}]/H$\alpha) < -0.5$; 
3) photo-ionized zone: $\log($[S {\footnotesize II}]/H$\alpha) < -1$. 
}\label{fig:theorysiishock}
\end{figure}

\begin{figure}[t]
\centering
\includegraphics[height=4.0 in]{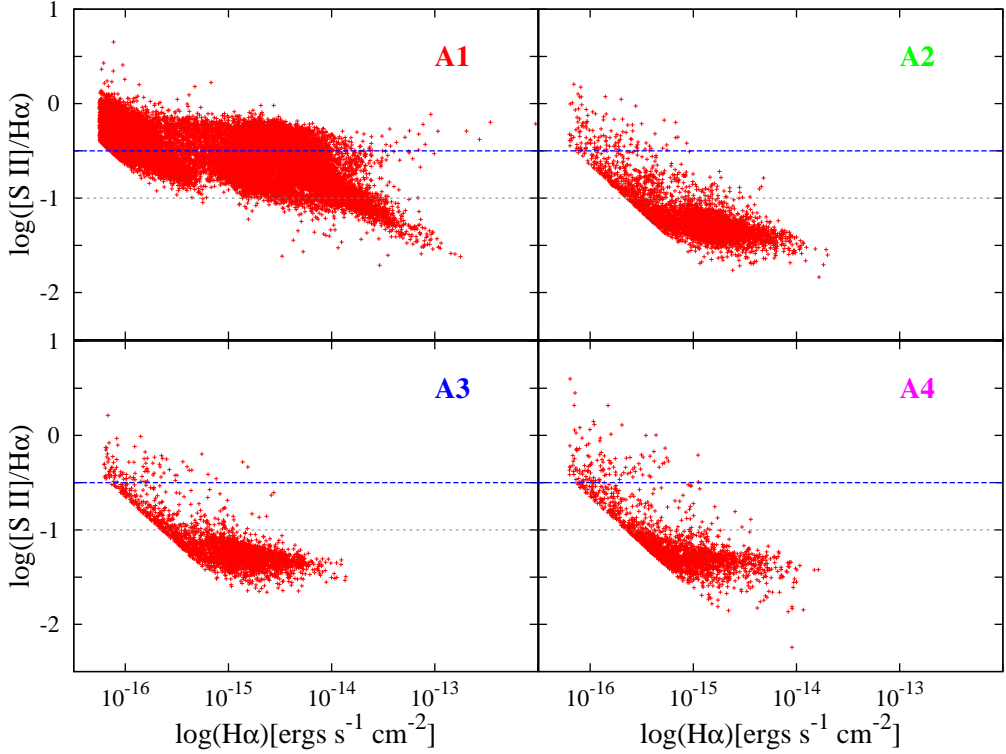}
\caption{The [S {\footnotesize II}]/\ha ratio vs. the dust extinction corrected \ha flux for the pixels in A1, A2, A3, and A4. 
The vertical and oblique cuts to the left and bottom-left of the data points are our $4~\sigma$ detection limits. 
The $1~\sigma$ limit for each line emission is summarized in Table 1. 
Most data points in A2, A3, and A4, have the ratios lower than -1; the photo-ionized zone in Figure~\ref{fig:theorysiishock}. 
Similarly to what found in the diagnostic diagram of Figure~\ref{fig:diag4regions}, 
the regions in the spiral arm have little shock-ionized gas. In A1, most data points reside in the mixing zone, 
$-1 < $ \sii/\ha $ < -0.5$. 
}\label{fig:obssiishock}
\end{figure}

Finally, as  the low surface brightness of \oiii$~$ is the major limiting factor for our diagnostics, 
we attempt to use the single line ratio criterion, [S {\footnotesize II}]/\ha $> -0.5$, to overcome this limitation (M4 in Table 2).   
This is another generous criterion for discriminating shocks from photo-ionization, like the M3 criterion, 
although it provides a less accurate separation than the diagnostic diagram involving the [OIII] line; 
however, it does not suffer from the limitations produced by faint or undetected \oiii. 
Figure~\ref{fig:theorysiishock} shows the theoretical relation between normalized \ha flux and [S {\footnotesize II}]/\ha 
ratio from the continuous starburst model of K01 and the shock+precursor model of A08. 
In this figure, we define three regions: (1) the shock zone, $\log($[S {\footnotesize II}]/\ha)$ > -0.5$, (2) the mixing zone, $-1 < \log($[S {\footnotesize II}]/\ha $) < -0.5$, and (3) the photo-ionization zone, $\log($[S {\footnotesize II}]/\ha$) < -1$. 
Our single-ratio diagnostics, $\log($[S {\footnotesize II}]/\ha$) > -0.5$, 
thus contains contributions from photo-ionized gas at low surface brightness. 
We consider this ``contamination'' acceptable, in view of the fact that 
we recover all signal from shocks. Figure~\ref{fig:obssiishock} shows the observed relation between 
the dust extinction corrected \ha flux and the [S {\footnotesize II}]/\ha ratios for A1, A2, A3, and A4. 
The comparison of Figures~\ref{fig:theorysiishock} and~\ref{fig:obssiishock} 
suggests that shocks are present in all four regions analyzed. 

\begin{figure}[t]
\centering
\includegraphics[height=4.8 in]{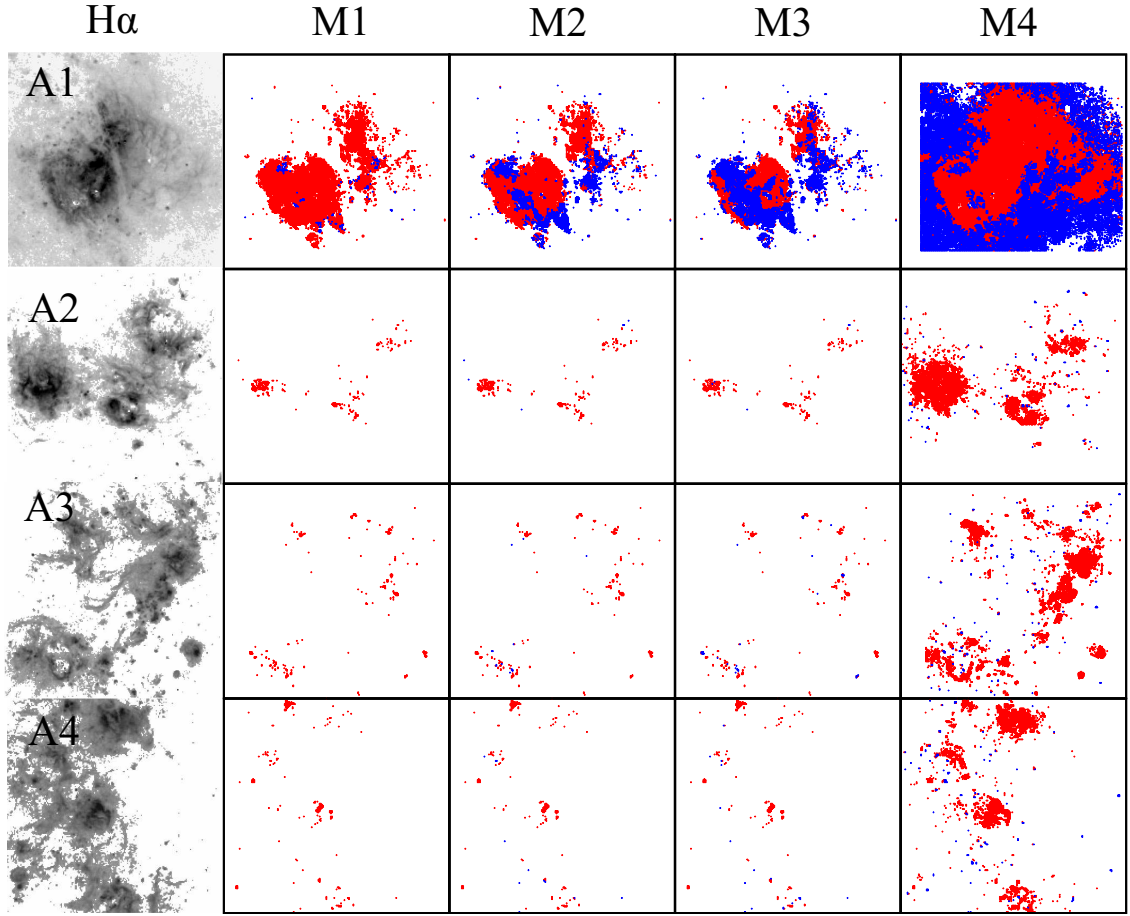}
\caption{The \ha images (4 $\sigma$ detection limit) covering the regions, A1, A2, A3, and A4 (first column), 
the locations of photo-ionized gas (red) and shock-ionized gas (blue) with 
the shock criteria M1(second column), M2(third column), M3(forth column), and M4(fifth column). 
In the M4 criterion, some faint diffuse photo-ionized gas with low ionization parameter is counted as shock-ionized gas, 
thus, producing contamination in the shocked gas and causing an overestimate of both the luminosity 
and area of the shock component. 
}\label{fig:shockareas}
\end{figure}

In Table 2, the SFR is calculated from the photo-ionized gas only; i.e. (total \ha luminosity - shock \ha luminosity), 
corrected for the effects of extinction using the WFC3 P$\beta$($\lambda 1.282 \mu m$) image (Liu et al. 2011, in prep).  
The SFR density is calculated by dividing the SFR by the area of the photo-ionized pixels. 
Figure~\ref{fig:shockareas} shows the spatial distribution of the ionized gas from Table 2 
according to each shock separation criterion, M1, M2, M3, and M4; 
the \ha images (4 $\sigma$ cut) covering the regions, A1, A2, A3, and A4, are shown 
in the first column together with the locations of photo-ionized gas (red) and shock-ionized gas (blue) with the shock criteria of M1(the second column), M2(the third column), M3(the fourth column), and M4(the fifth column). 

In the spiral arm (regions A2, A3, A4), the shock gas from the ``Maximum Starburst Line'' criteria(M1) is negligible. 
In the other criteria(M2, M3, and M4), the luminosity ratios and the area ratios are 4\% or less. 
Figure~\ref{fig:obssiishock} also shows that most of the $\log($[S {\footnotesize II}]/H$_{\alpha})$ values in A2, A3, and A4, are less than -1, 
so located in the photo-ionization zone. Therefore, feedback from star formation is very low in the spiral arm of M83 
perhaps owing to the youth of the H {\footnotesize II} regions we detect. 
Shocks remain a small or negligible component in the outer H {\footnotesize II} regions.  

For the central starburst, the SFR is 2.7 times larger than the previous study of C04 
due to  a better \ha filter coverage in the present study 
(the WFPC2 narrow-band filter used in the C04 study placed the redshifted \ha on the red shoulder of the filter).
In addition, the P$\beta$/H$\alpha$ extinction map we use probes larger dust extinction values than the H$\alpha$/H$\beta$ extinction map (Liu et al. 2011; in prep). 
The SFR density of A1 is 3 times larger than the values of A2, A3, and A4.
For the conservative shock estimate(M1), the luminosity ratio and the area ratio of A1 are consistent with the previous study of C04 when we consider the 
range of values for different [N {\footnotesize II}] corrections of the \ha filter. In this conservative criterion, a few percents of the \ha luminosity comes from shock gas. 
The area occupied by the shock gas is larger than the luminosity contribution, consistent with the fact that the shock gas has lower \ha brightness  but is 
distributed more diffusely than the photo-ionized gas. 
For the other criteria (M2, M3, and M4), the shock \ha luminosities can reach 15\% to 33\% of the 
total luminosity. The covering areas of shocks are also larger totaling 28\% to 53\% of the total area. 
The M3 and M4 criteria are generous, and may misclassify as shocks some regions that could actually 
contain mostly photo-ionized gas; we thus expect the shock luminosity and area can be overestimated for 
M3 and M4.
We conclude that the acceptable values are 15\% for the \ha luminosity and 30\% for the areal coverage of shocks in the center of M83. 
When comparing our 15\% fraction of shocked gas in the central starburst of M83 with expectations for the mechanical energy 
output from the starburst itself (4\% to 9\%, columns 4 and 8 of Table 6 in C04), we conclude that virtually all of the available mechanical energy 
from massive star winds and supernovae is radiated. This is in agreement with the findings of Martin (2006), who conclude 
that only a few percent of the mechanical energy from star formation is available as kinetic energy to drive the ISM.

\subsection{Shock-ionized Gas}
In this section, we compare our data with shock-ionization models and investigate the properties of the shock-ionized gas. 
The shock-ionization model of A08 uses the ionizing radiation field from hot radiative shock layers  dominated by free-free emission 
and the MAPPINGS III code to calculate the gas ionization state and the intensity of the emission lines. 
The emission from shocks consists of two components; the shock layer (postshock component) and the precursor (preshock component). 
The shock layer is the cooling zone of the radiative shock and the precursor is the region ionized by the upstreaming photons from the cooling zone.  
Since free-free emission is dominant in shock layers,  the ionizing radiation field from shocks is mainly determined by 
the shock velocity and the preshock gas density (see A08 for more details). 
So for a given metallicity and pre--shock gas density, 
the shock velocity is the main parameter determining the line ratios in the diagnostics. 
The main branch of the shock grids labeled with ``B=1.0E-4'' in Figure~\ref{fig:diag} shows the line ratios 
for shock velocities from 100 km/s to 900km/s, 
with a fixed ISM magnetic field of $10^{-4} \mu G$ . 
The magnetic field in the ISM plays an important role in determining 
the post-shock gas density: higher ISM magnetic fields correspond to 
lower post-shock densities, with major effects on the ionization 
parameter of the post-shock gas component.
Hence, the magnetic field is the second main parameter in shock models. The side branches from the selected shock velocities, 100, 150, ..., 500 km/s, show the 
effect of changing the magnetic fields from $10^{-4}\mu G$ to $10\mu G$. The metallicity of the model is 2$Z_{\odot}$ and the preshock gas density is 1 $cm^{-3}$.  

Two peculiar features can be identified in Figure~\ref{fig:diag}, 
in the region dominated by shock-excitation. One is 
a series of pixels in a roughly horizontal sequence, 
around $\log$([O {\footnotesize III}]/\hb) $\sim 0$, which seem to be consistent with 
the horizontal feature of the shock+precursor model from A08. 
This horizontal distribution of shock-ionized gas is a unique feature 
when compared with other solar and sub-solar abundant galaxies 
such as NGC3077, NGC4214, and NGC5253 in C04, which have the shock-ionized gas 
near or outside of the ``Maximum Starburst Line". 
A part of the feature is marked by a box labeled `S2' in the figure, 
where it is outside of the photo-ionization locus. So, the `S2' region can be considered as a  
shock-ionized region with high shock velocities ($> 300$ km/s) according to the shock models. 
The other peculiar feature is a vertical plume located at $\log$([O {\footnotesize III}]/\hb) $> 0.2$ 
and $-1 < \log$([S {\footnotesize II}]/\ha) $< -0.5$, 
where some low velocity shocks are likely to be present based on  
the shock models. This region, also marked 
by a box in Figure~\ref{fig:diag}, is labeled `S1'. 
Both regions belong to the nuclear region A1, where the most active star formation in 
the galaxy is taking place. 

\begin{figure}[t]
\centering
\includegraphics[height=4.5 in]{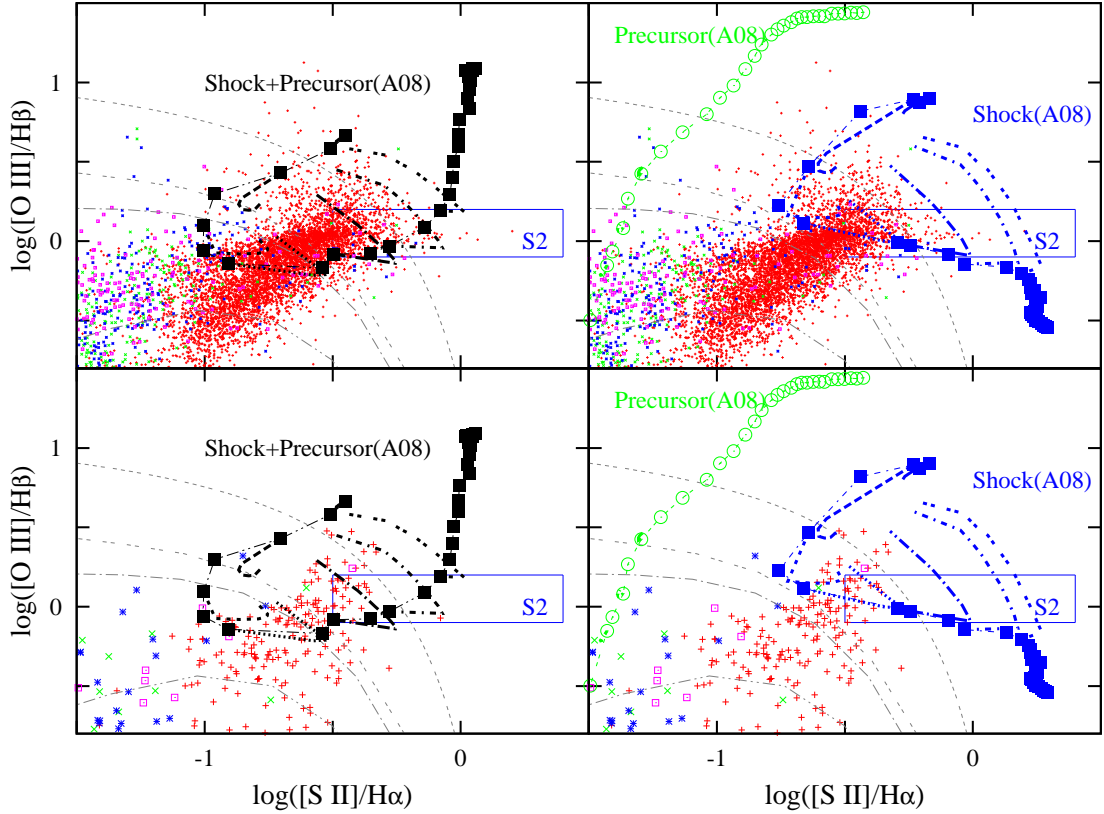}
\caption{Top: The Shock+Precursor model of A08 (left) and the Shock-Only model (blue) with the Precursor-Only model (green) of A08 (right). Bottom: The rebinned points by $35\times35$ pixels square corresponding to $1.4''$ with the shock+precursor model of A08 (left) and the Shock-Only model (blue) with the Precursor-Only model (green) of A08 (right).
}\label{fig:shockzoom}
\end{figure}

\begin{figure}[t]
\centering
\includegraphics[height=7.0 in]{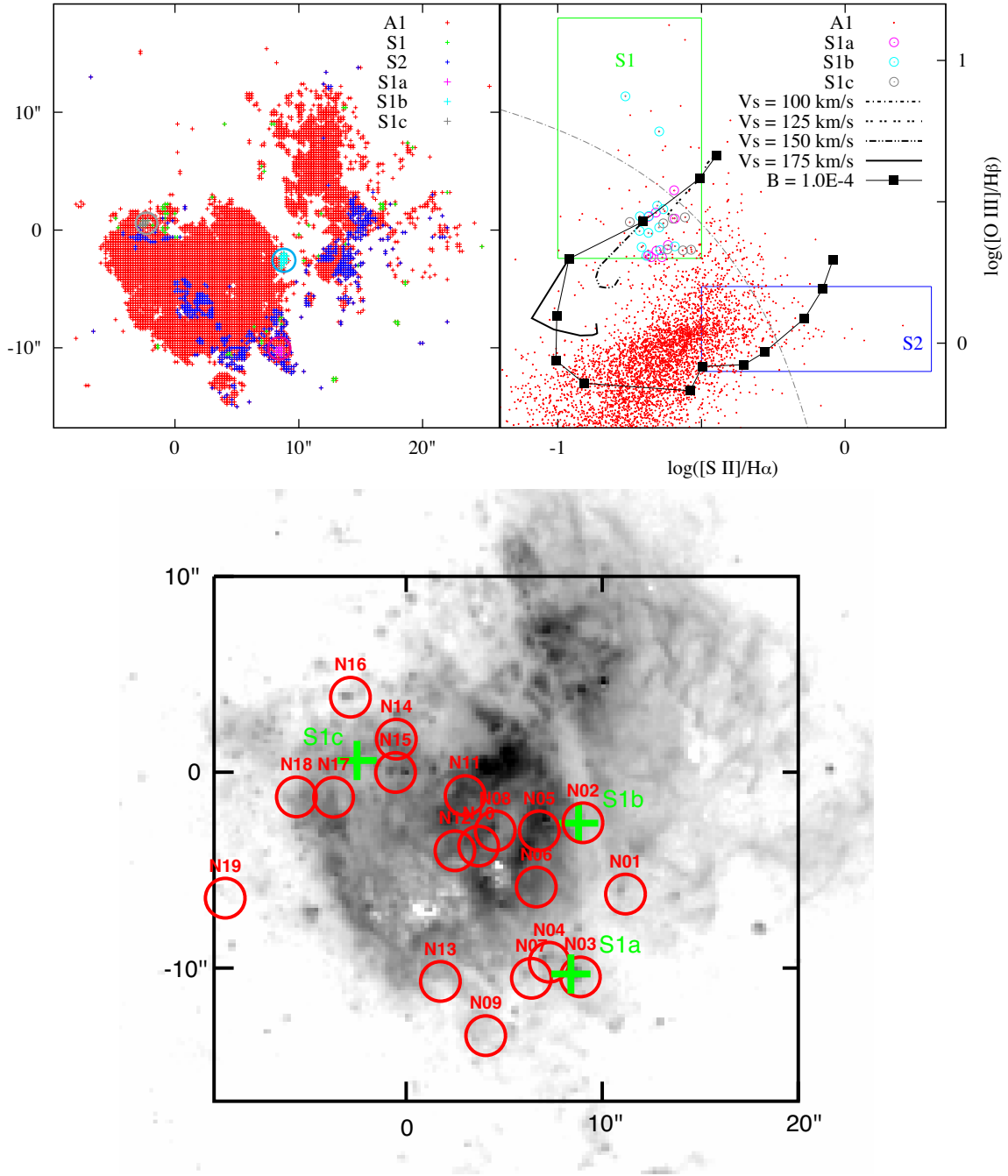}
\caption{Top left : The physical locations of the ``S1''(green point) and the ``S2''(blue point) in A1. 
Many S1 pixels seem to be distributed randomly. Those can be noises because of $4\sigma$ cuts. 
But some pixels show geometrical correlation and we locate and recolor the spots as S1a(magenta), S1b(cyan), and S1c(grey) with open circles. Top right : The identifications of S1a, S1b, and S1c in the diagnostic diagram. The S1a, S1b, and S1c follow the low velocity shock grids well. Bottom: The positions of SNRs in Dopital et al. (2010) in red circles and our low velocity shock components, S1a, S1b, and S1c, in green crosses. 
}\label{fig:central}
\end{figure}

The horizontal feature around S2 in the diagnostic diagram is highlighted 
in Figure~\ref{fig:shockzoom}, where the data are compared with:
the shock+precursor models from A08 (top-left panel), and the precursor 
and the shock shown as separate curves, also from A08 (top-right 
panel). To assess again the impact of spatial sampling 
on our data, we reproduce the top panel with the data rebinned 
in boxes of 35x35 pixels, or $1.4''\times1.4''$(bottom panels). The models clearly indicate 
somewhat different natures for the horizontal feature. The A08 
models suggest that the feature could be due to shocks-only or shock+precursor 
(but not precursor-only model), with velocities 
in the range ~250-350 km/s. 
The pixels corresponding to the S2 data points in the diagnostic diagram of Figure~\ref{fig:central} are 
re-projected onto a spatial reproduction of A1 in the left panel of 
Figure~\ref{fig:central}. Those points are mostly located at the edges of the 
region of photo-ionization, where we would expect to find shock-ionized 
gas. In addition, many S2 points are located inside the largest of 
the regions in A1, in correspondence of the `bubble' created by 
one or more of the young stellar clusters in the starburst (Harris et al.,
2001); this is also an area where we do expect shocks 
to be present. 

Region S1 in Figure~\ref{fig:diag} is enlarged in the right panel of Figure~\ref{fig:central}, 
and its pixels spatially projected in the left panel of the same 
figure. A number of those data points from the diagnostic 
diagram, especially those with the highest $\log$([O {\footnotesize III}]/\hb), 
distribute randomly on the spatial projection, indicating that 
they are the result of noisy pixels combined with our generous 
4~sigma cut in the line emission images. However, a number of pixels 
in the S1 region appear correlated, and, like most of the pixels 
from S2, crowded in regions at the edges of the A1 line emission area. 
We name those regions S1a(magenta), S1b(cyan), and S1c(grey), 
indicated with open circles in the left panel, and 
identify those regions in the diagnostic diagram in the right panel of Figure~\ref{fig:central}.
Those data points, marked with circles in the right panel of the figure, 
are consistent with low velocity shocks+precursor models by A08, 
with velocity in the range ~100--150~km/s. These areas (also marked 
with open circles in the left panel of Figure~\ref{fig:central}) are located 
close to identified supernova remnants (SNRs) from Dopita et al. (2010) 
as shown in the bottom of Figure~\ref{fig:central}. 
Using the naming convention from that paper, we find SNRs N3,N4, and N7 
close to our region S1a, N2 to S1b, and N14, N15, N17, and N18 to S1c. 
S1a and S2b seem to correspond to the SNRs, N3 and N2, respectively. 
There is no corresponding SNR for S1c. It is interesting that S1c is surrounded by 
the four SNRs, N14, N15, N17, and N18. However, 4 SNRs, N14, N15, N17 and N18 
are located in correspondence of feature both in S1 and S2 (top left with bottom panel of Figure~\ref{fig:central}). 
Thus, it is likely that the identified low-velocity shocks are closely related to the identified SNRs. 

Two of the S1 subregions (S1a and S1b)
are located adjacent to the dust lane that crosses the center of 
M83. Although the diagnostic diagram tends to be fairly 
unaffected by dust extinction, the large dust column densities found 
in dust lanes may have an impact on the line ratios. We can not  
eliminate this possibility with the data in hand. Finally, dynamical 
interactions may play a role in the creation of low velocity shocks. 
S1a and S1c are located at the junction points of the spiral arms to 
the central outer ring, while S1b is in--between the spiral arm and 
the outer ring. The interaction between those morphological 
features could be at the basis of the observed shocks. 

The width of the precursor of a shock front is generally  larger than 
our binned pixel size($\sim 4.3$pc), while photo-ionization equilibrium 
can be achieved well within the size. For the shock model of A08 with solar abundance, 
1 $cm^{-3}$ preshock gas density, 200 km/s shock velocity, and 3.2 $\mu$G magnetic field, 
the length of preshock ionized zone is 40 pc, while the length of postshock cooling zone is less than 4 pc. 
Even though the postshock zone (shock layer) can be small enough to fit in 
our bin size, the preshock component is much larger than our bin scale. 
This means that , unlike the photo-ionization model, 
the shock-ionization model needs to be applied carefully when we compare it 
with the data. For instance, if a pixel line-of-sight passes through 
both of precursor and shock front, the pixel line ratio is meaningful 
when compared with the shock+precursor grids. 
If a pixel only covers a part of the precursor region or  shock front, 
the comparison has to be made with shock-only or precursor-only models. 
When we consider the complicated geometry through observed line-of-sights, 
it is hard to determine which light-of-sight is for shock-only, precursor-only or shock+precursor. 
Therefore, we may need to make a geometrical block enough 
to cover both of shock and precursor, but excluding photo-ionized gas, 
and sum up the whole line flux in the block in order to compare it 
with the shock+precursor model. So basically shock-ionization models may need, 
in principle, to be compared on a large-scale basis than what done so far. 
However, as shown in the bottom panels of both Figures ~\ref{fig:photozoom} and \ref{fig:shockzoom}, 
rebinned to pixel sizes of 1.4$''$ ($\sim$30~pc), the quantitative 
distribution of the data points on the diagnostic diagram does 
not change, implying that our smaller pixel scale is adequate to 
average out effects of spatial variations in shocks.

\section{Summary}

We have presented emission line diagnostic diagrams 
for the central $\sim$4~kpc of the nearby starburst galaxy M83, 
using the newly available WFC3 images 
in order to investigate the properties and morphology of both 
the photo-ionized and shock-ionized gas on a pixel-by-pixel basis.  
Here is the summary of our results: 

1. A comparison of the data with a number of photo- and shock-ionization 
models show that the best fit for the data of the central starburst region 
(A1 in our convention) is given by a combination of the Z=2 Z$_{\odot}$ photo--ionization 
models of K01 and the shock+precursor models of A08. 

2. Regions at increasing galactocentric distance show an increasing 
[O {\footnotesize III}]/\hb ratio for the photo--ionized gas, consistent with 
the presence of a metallicity gradient in the galaxy. 

3. Changing the discriminating `line' between shocks and photo--ionized 
gas has major implications for the fraction of ionized gas that can be 
attributed to shocks. The Maximum Starburst Line of K01 
(the most conservative criterion for identifying shocks; See Kauffmann et al. 2003) 
produces fractions of a 
few percent, in agreement with previous results (e.g., C04). Adopting 
a metallicity--dependent criterion, the fraction of shock--ionized 
H$\alpha$ luminosity increases to about 15\%, 
and is in any case no larger than about 33\% in the starburst center, 
while shocks are always negligible in the spiral area H {\footnotesize II} regions. 
At the 15\% level, this requires that virtually all the mechanical energy 
produced in the starburst center is radiated away in the shocks. 

4. We find three regions dominated by low velocity shocks. All three 
are located at the edges of the central starburst, and can have 
one of three possible origins: a) can be the result of feedback from 
SNRs located near each component; b) can be due to the dynamical 
interaction between the center and the spiral arm; c) could be the 
result of the extreme local dust extinction. Discriminating among these 
three scenarios will require additional more sophisticated observations.

\acknowledgments
We are grateful to an anonymous referee for comments that have improved this 
paper. This paper is based on observations taken with the NASA/ESA Hubble Space Telescope obtained at the Space Telescope Scinece Institute, which is operated by AURA, Inc., under NASA contract NAS5-26555. It uses Early Release Science observations made by the WFC3 Science Oversight Committee. We are grateful to the Director of STScI for awarding DirectorÕs Discretionary time for this program. 


\begin{deluxetable}{lrccc}
\tabletypesize{\footnotesize}
\tablecolumns{4}
\tablewidth{0pc}
\tablecaption{WFC3 Observation (Program ID 11360)}
\tablehead{
\colhead{Filter}& \colhead{Exposure (sec)} & \colhead{Date Obs. (2009)} & \colhead{Continuum\tablenotemark{a}} & \colhead{$1~ \sigma$ Limit \tablenotemark{b}}
}
\startdata
F487N (H${\beta}$) &  $3\times 900$ & Aug 25  & F555W & $4.3 \times 10^{-18}$ \\
F502N (\oiii) & $3\times 828$   &  Aug 26 & F555W & $4.6 \times 10^{-18}$ \\
F657N (H$\alpha$) &  $4\times 371$  &  Aug 25 & 0.58 F555W + 0.42 F814W & $2.8 \times 10^{-17}$ \\
F673N (\sii) & $2\times 600, 650$  &  Aug 20 & 0.58 F555W + 0.42 F814W & $5.8 \times 10^{-18}$ \\
F555W & $3\times 401$   &  Aug 26 & ... & ... \\
F814W &  $3\times 401$  &  Aug 26 & ... & ... \\
\enddata
\tablenotetext{a}{The continuum images used for subtracting stellar continuums from narrowband images. The interpolated wavelength is 6600\AA (660 nm) for both of [S {\footnotesize II}] and \ha. Because the filter widths of the broad band images are much larger than the wavelength separation between our chosen wavelength (660nm) and each of the red emission lines, our rough single interpolation is acceptable with typical errors of less than 5\%.}
\tablenotetext{b}{The 1 $\sigma$ detection limit for each emission-line, in units of ergs s$^{-1}$ cm$^{-2}$ per bin, 
where a bin is $5 \times 5$ pixels (or $0.2''\times 0.2''$), after subtraction of the stellar continuum.}
\end{deluxetable}

\begin{deluxetable}{ccrrrrrr}
\tabletypesize{\tiny}
\tablecolumns{8}
\tablewidth{0pc}
\tablecaption{Measured and derived quantities}
\tablehead{
\colhead{Shock Separation Criteria}& \colhead{Region} & \colhead{$L_{H_\alpha,tot}$ \tablenotemark{a}} & \colhead{$L_{H_\alpha,sh}$\tablenotemark{b}} & \colhead{SFR\tablenotemark{c}} & \colhead{SFR density\tablenotemark{d}} & \colhead{ $L_{H_\alpha,sh}/L_{H_\alpha,tot}$\tablenotemark{e}} & \colhead{ $A_{sh}/A_{tot}$\tablenotemark{f}} \\
\colhead{(Separation Model Name)} &\colhead{}  & \colhead{ (erg s$^{-1}$)} &\colhead{(erg s$^{-1}$)} & \colhead{($M_{\odot}$ yr$^{-1}$)}  & \colhead{($M_{\odot}$ yr$^{-1}$ kpc$^{-2}$)} &\colhead{} &\colhead{} 
}
\startdata
Maximum Starburst Line  & A1 & $1.07\times 10^{41}$ & $1.71\times 10^{39}$ & 0.831  & 8.30 &  0.016 (0.030, 0.012) &  0.029 (0.16, 0.07)\\
(M1)  & A2 & $1.36\times 10^{39}$    &   0.0  &   0.011    &   2.55     &  0.0  &  0.0\\
  & A3 & $1.53\times 10^{39}$    &   0.0   &  0.012   &    2.70   &    0.0    &  0.0\\
  & A4 & $1.04\times 10^{39}$   &    0.0  &  0.008    &   2.18    &  0.0  &    0.0\\
  \cline{1-8}\\
N350 continuous Z1.0 Line  & A1  &$1.07\times 10^{41}$ &  $1.57\times 10^{40}$    &  0.720  &     9.77  &    0.147   &   0.284\\
(M2) & A2  & $1.36\times 10^{39}$ &  $1.56\times 10^{37}$   &  0.011    &   2.59   &  0.012   &  0.026 \\
 & A3  & $1.53\times 10^{39}$  & $1.61\times 10^{37}$   &  0.012    &   2.83   &  0.010   &  0.058 \\
 & A4  &  $1.04\times 10^{39}$ &  $8.01\times 10^{36}$ &   0.008   &    2.29  &  0.008   &  0.054\\
   \cline{1-8}\\
Shock Locus  & A1  &$1.07\times 10^{41}$ &  $3.49\times 10^{40}$    &  0.569  &     12.0  &    0.326   &   0.539\\
(M3) & A2  & $1.36\times 10^{39}$ &  $3.47\times 10^{36}$   &  0.011    &   2.60   &  0.003   &  0.022 \\
 & A3  & $1.53\times 10^{39}$  & $6.43\times 10^{37}$   &  0.012    &   3.00   &  0.042   &  0.140 \\
 & A4  &  $1.04\times 10^{39}$ &  $1.36\times 10^{37}$ &   0.008   &    2.36  &  0.013   &  0.089\\
   \cline{1-8}\\
$\log([$S II$]/$H$_\alpha) > -0.5$  & A1  & $2.09\times 10^{41}$ &  $4.13\times 10^{40}$    &   1.324  &    5.64   &   0.198  &    0.500\\
(M4) & A2  & $1.08\times 10^{40}$ &  $3.94\times 10^{37}$   &  0.085   &   1.56 &  0.004  &   0.031\\
 & A3  & $1.53\times 10^{40}$  & $5.25\times 10^{37}$    &  0.121     &  1.69  &  0.003   &  0.032 \\
 & A4  & $6.31\times 10^{39}$  & $5.13\times 10^{37}$   &  0.049   &    1.24  &  0.008  &   0.054\\
\enddata
\tablenotetext{a}{Total \ha luminosity in each region.}
\tablenotetext{b}{\ha luminosity of shocked gas.} 
\tablenotetext{c}{Star formation rates from the \ha flux of the photo-ionized gas corrected for extinction (See section 3.2).}
\tablenotetext{d}{SFR divided by the pixel areas (Average SFR density per pixel area).}
\tablenotetext{e}{Fraction of the total \ha luminosity associated with shock-ionized gas.}
\tablenotetext{f}{Fraction of the H$\alpha$ area occupied by shock-ionized gas. The values in parenthesis of `e' and `f' columns are from C04 and correspond to different [N {\tiny II}] corrections. See C04 for more details.}
\end{deluxetable}

\end{document}